\documentclass[onecolumn]{PAZH_new}

\usepackage{graphicx}
\usepackage{latexsym,amssymb,amsmath,color}
\usepackage{hyperref}

\begin{document}
\onecolumn

%%% Список авторов задается командой \authors.
%%% В квадратных скобках размещается краткий список для верхнего колонтитула на четных страницах:
%%% ФАМИЛИЯ (заглавными буквами) - если автор только один;
%%% ФАМИЛИЯ1, ФАМИЛИЯ2 - если авторов двое;
%%% ФАМИЛИЯ1 и др. - если число авторов больше двух.
%%% В фигурных скобках размещается полный список авторов,
%%% причем каждый следующий добавляется командой \nextauth, имеющей два обязательных аргумента:
%%% в первых фигурных скобках указываются инициалы и фамилия, а во вторых - номера, соответствующие аффилиациям.
%%% Дополнительно, для основного соавтора в квадратных скобках перед фигурными указывается электронный адрес.

\authors[MARKOZOV et al.]{
  \nextauth[markozoviv@mail.ru]{I. D. Markozov}{1,2},
  \nextauth{G. K. Pimenov}{3},
  \nextauth{A. A. Mushtukov}{4}
}

%%% Полный и краткий заголовки статьи задаются командой \titles.
%%% В квадратных скобках размещается краткий заголовок для верхнего колонтитула на нечетных страницах
%%% (для корректной печати этого колонтитула команда \titles должна располагаться после команды \authors).
%%% В фигурных скобках дается полное название.

\titles[Gravitational Waves from Her X-1]{Modeling the Gravitational Wave Signal from the X-ray Pulsar Her X-1}

%%% Список аффилиаций задается командой \affiliations, внутри которой каждая следующая аффилиация задается командой \nextaffil. 
%%% При этом последовательная нумерация аффилиаций производится автоматически.
%%% Однако, авторы должны сами проконтролировать соответствие этому списку номеров, указанных в командах \nextauth выше.

\affiliations{
\nextaffil{Ioffe Institute, Russian Academy of Sciences, St. Petersburg, 194021 Russia}
\nextaffil{Space Research Institute, Russian Academy of Sciences, Moscow, 117997 Russia}
\nextaffil{St. Petersburg State University, St. Petersburg, 199034 Russia}
\nextaffil{Mullard Space Science Laboratory, University College London, London, UK}
%\nextaffil{Astrophysics, Department of Physics, University of Oxford, Oxford, UK}
}

%%% Аннотация задается как аргумант команды \wideabstract. 
%%% В конце аннотации, внутри этой команды, задаются ключевые слова как аргумент команды \keywords.
%%% После них можно вставить "\doi{}", чтобы зарезервировать место для DOI.

\wideabstract{
 The paper is devoted to modeling the gravitational waves from the X-ray pulsar Her X-1. The neutron star is considered as a freely precessing ellipsoid. The gravitational wave signal is calculated in the quadrupole approximation. The $h_{\times}$ and $h_{+}$ polarization profiles are constructed for the precession parameters of the neutron star in Her X-1 measured in up-to-date works. The possibility of signal detection with the DECIGO telescope is assessed. The formulas for the semi-analytical modeling of the gravitational wave signal from precessing neutron stars by the method of perturbations in the small neutron star ellipticity parameter are deduced.

\keywords{neutron stars, X-ray sources, gravitational waves}

\doi{10.1134/S106377372670026X}
} 

%=========================================================

\section*{Introduction}
\label{introduction}
X-ray pulsars are neutron stars (NSs) in binary systems in which accretion from the companion star onto the NS occurs. As a result of accretion, the polar regions of the neutron star shine in X-rays, with the X-ray light curve of such an object containing pulsations due to the misalignment of the NS rotation and magnetic axes. The pulsation periods for such objects range from $\sim0.1$~s to several hundred seconds (Mushtukov and Tsygankov 2022).

As a result of its rapid rotation, a NS that is not an ideal sphere must be subject to precession. Precessing NSs have been recognized as potential sources of continuous gravitational waves (GWs) for decades (Jones and Andersson 2002). In the works of Zimmermann and Szedenits (1979, 1980), which underwent further development in Van Den Broeck (2005) and Gao et al. (2020), the GW signal from a NS was modeled in the quadrupole approximation, with the NS having been considered as an absolutely rigid ellipsoid.
 
The results of theoretical studies suggested that the GW signal from precessing millisecond pulsars could be detected with ground-based GW telescopes of the existing or next generation. However, by now the search for such a signal has not yielded a positive result (Abac et al. 2025).

The NS to be discussed in this paper is a member of the binary system Her X-1 and is observed as an X-ray pulsar. The object is 7~kpc away from us and has a pulsation period of 1.24~s and an orbital period of 1.7~days (Tananbaum et al. 1972; Cherepashchuk et al. 1972). The binary system also exhibits a superorbital period of 35~days (Giacconi et al. 1973). Owing to the wealth of observational data, including X-ray polarimetry (Doroshenko et al. 2022), the rotation parameters of the NS in Her X-1 and the parameters of the accretion onto it are known, by the standards of the accuracy of measurements for X-ray pulsars, fairly well.
 
One of the possible explanations for the superorbital period of 35~days includes free NS precession (Brecher 1972; Novikov 1973). The hypothesis of a NS precessing with a period of 35~days in the pulsar Her X-1 allows one to explain the up-to-date multiwavelength observational data (Postnov et al. 2013; Kolesnikov et al. 2022) and the X-ray polarization measurements for Her X-1 with the IXPE telescope (Heyl et al. 2024). An alternative to it suggests the precession of a warped accretion disk (see, e.g., Peterson 1977; Leahy and Frost 2025) and does not deal with the NS precession.
 
In this paper we investigate the continuous GWs from the pulsar Her X-1 that must result from the NS precession in this binary system. We model the GW signal in the quadrupole approximation (see Landau and Lifshitz 2003) by approximately considering the NS as an ellipsoid. Based on the derived profiles of the signal for the $+$ and $\times$ polarizations, we discuss the possibility of detecting these waves with the DECi-hertz Interferometer Gravitational wave Observatory (DECIGO) telescope.

\section*{Neutron Star Precession}
\label{Precession}

We consider the NS either as a precessing triaxial ellipsoid or as a prolate axisymmetric ellipsoid, which is a special case of the triaxial one. We assume the rotation to be rigid and the precession to be free, i.e., all torques to be compensated. The model of a freely precessing uniaxial ellipsoid and the model of a triaxial ellipsoid were used to explain the observational data in Postnov et al. (2013) and Kolesnikov et al. (2022), respectively. Both these models were considered when explaining the polarization data in Heyl et al. (2024).

The precession is described by the Euler angles $\phi,\: \psi,\: \theta$, where $\phi$ is the precession angle, $\psi$ is the rotation angle, and $\theta$ is the wobble angle. The letter designations $\phi,\: \psi,\: \theta$ were introduced in Section 35 of the book by Landau and Lifshitz (2004).

Let us also introduce the angular velocity vector $(\omega_1,\omega_2,\omega_3)$ whose components are related to the Euler angles as follows:
 \begin{equation}
 \label{eq:omeg1}
 \omega_1 = \dot{\phi}\sin\theta\sin\psi+\dot{\theta}\cos\psi
 \end{equation}
 \begin{equation}
 \label{eq:omeg2}
 \omega_2 = \dot{\psi}\sin\theta\cos\psi-\dot{\theta}\sin\psi
 \end{equation}
 \begin{equation}
 \label{eq:omeg3}
 \omega_3 = \dot{\phi}\cos\theta+\dot\psi
 \end{equation}

The evolution of the angular velocity with time is described by the Euler dynamical equations
 \begin{equation}
 \label{eq:euler}
 \begin{cases}
  I_1 \dot{w_1} - (I_2-I_3)w_2w_3 = 0\\
  I_2\dot{w_2} - (I_3-I_1)w_3w_1 = 0\\
  I_3\dot{w_3} - (I_1-I_2)w_1w_2 = 0,
 \end{cases}
\end{equation}
where $I_i$ are the principal moments of inertia and $\omega_i$ are the components of the angular velocity vector. To completely describe the precession, we need to know $\omega_i$ and the Euler angles at each instant of time. The angular velocities are obtained by integrating the system of equations (\ref{eq:euler}). This system has the following integrals of motion: the angular momentum $J=\sqrt{I_1^2w_1^2+I_2^2w_2^2+I_3^2w_3^2}$ and the kinetic energy $E=(I_1w_1^2+I_2w_2^2+I_3w_3^2)/2$.

\section*{The Solving Technique}
\label{sol_method}

We relied on the technique for solving the Euler equations described in Betsch and Siebert (2009) and used to model the GWs from a precessing triaxial NS in Gao et al. (2020).
     
For the angles, as for the angular velocities, there exists a solution in quadratures, but its explicit calculation is time consuming. Therefore, we decided to obtain the values of these variables by numerically integrating the dynamical equations. For this purpose, we used the quaternion $q = q_0 + q_1\mathbf{i} + q_2\mathbf{j} + q_3\mathbf{k}$ whose components are related to the Euler angles as follows:

     \begin{equation}
     \label{eq:q0}
 q_0 = \cos{\frac{\theta}{2}}\cos{\left(\frac{1}{2}(\phi+\psi)\right)}
\end{equation}
\begin{equation}
\label{eq:q1}
 q_1 = \sin{\frac{\theta}{2}}\cos{\left(\frac{1}{2}(\phi-\psi)\right)}
\end{equation}
\begin{equation}
\label{eq:q2}
 q_2 = \sin{\frac{\theta}{2}}\sin{\left(\frac{1}{2}(\phi-\psi)\right)}
\end{equation}
\begin{equation}
\label{eq:q3}
 q_3 = \cos{\frac{\theta}{2}}\sin{\left(\frac{1}{2}(\phi+\psi)\right)}
\end{equation}
     
The rotation matrix $\mathcal{R}$ of the rotating body is expressed via this quaternion:
 \begin{equation}
    \begin{aligned}
     \mathcal{R}_{11}=q_0^2 + q_1^2 - q_2^2 -q_3^2, \quad \mathcal{R}_{12}=2q_1q_2-2q_0q_3,\quad  \mathcal{R}_{13}=2q_1q_3+2q_0q_2, \\ \mathcal{R}_{21}=2q_1q_2+2q_0q_3, \quad 
     \mathcal{R}_{22}=q_0^2 - q_1^2 + q_2^2 -q_3^2, \quad \mathcal{R}_{23}=2q_1q_3-2q_0q_1 \\
     \mathcal{R}_{31}=2q_1q_3-2q_0q_2, \quad \mathcal{R}_{32}=2q_2q_3+2q_0q_1, \quad
     \mathcal{R}_{33}=q_0^2 - q_1^2 - q_2^2 +q_3^2
 \label{eq:R_matrix}
 \end{aligned}
 \end{equation}
    
The differential equation that defines the evolution of the quaternion can be expressed as
\begin{equation}
 \frac{\text{d}q}{\text{d}t} = \frac{1}{2} \begin{pmatrix}
      0 & -\omega_1 & -\omega_2 & -\omega_3 \\
      \omega_1 & 0 & \omega_3 & -\omega_2 \\
      \omega_2 & -\omega_3 & 0 & \omega_1 \\
      \omega_3 & \omega_2 & -\omega_1 & 0
                                                                    \end{pmatrix}
                                                                    \begin{pmatrix}
      q_0 \\
      q_1 \\
      q_2 \\
      q_3
\end{pmatrix}
\label{eq:qevolution}
\end{equation}

Equations (\ref{eq:euler}) and (\ref{eq:qevolution}) are integrated simultaneously using the DifferentialEquations library of the Julia programming language (\href{https://docs.sciml.ai/DiffEqDocs/stable/}{https://docs.sciml.ai/DiffEqDocs/stable/}). As a method for solving the system of equations, we chose the L-stable Rosenbrock method; the relative accuracy of the solution was specified to be $10^{-7}$.

\section*{The GW Profile}
\label{plus_cross}

Within general relativity, the GWs are decomposed into the sum of two polarizations (see, e.g., Landau and Lifshitz 2003):
\begin{equation}
h^{TT}=h_+(\hat{e}_1\otimes\hat{e}_1-\hat{e}_2\otimes\hat{e}_2)+h_\times(\hat{e}_1\otimes\hat{e}_2+\hat{e}_2\otimes\hat{e}_1).
\label{eq:summ_pol}
\end{equation}

Here $h^{TT}$ is the GW signal in the transverse--traceless gauge, $\hat{e}_1$ and $\hat{e}_2$ are two unit vectors lying in a plane perpendicular to the wave propagation direction. Suppose that the observer is in the $YZ$ plane, where the $Z$ axis is aligned with the direction of the conserved angular momentum of a freely precessing NS. In that case, if $i$ is the angle between the conserved angular momentum vector of the body and the line of sight, then the profiles of the GW signal from the two polarizations can be calculated using the following formulas (Gao et al. 2020):
\begin{equation}
h_{+} = -\frac{G}{rc^4}[(\mathcal{R}_{2k}\cos{i}+\mathcal{R}_{3k}\sin{i})(\mathcal{R}_{2l}\cos{i} + \mathcal{R}_{3l}\sin{i})-\mathcal{R}_{1k}\mathcal{R}_{1l}]A_{kl}
\label{eq:h_pl}
\end{equation}
\begin{equation}
 h_{\times} = -\frac{2G}{rc^4}(\mathcal{R}_{2k}\cos{i} + \mathcal{R}_{3k}\sin{i})\mathcal{R}_{1l}A_{kl}
 \label{eq:h_cr}
\end{equation}
where $G$, $c$, and $r$ are the gravitational constant, the speed of light, and the distance to the NS, respectively, and $\mathcal{R}$ is the rotation matrix expressed via the quaternions using Eqs. (\ref{eq:R_matrix}). The elements $A_{11}$ and $A_{12}$ of the matrix $A$ are represented by the formulas (\ref{eq:A11}) and (\ref{eq:A12}):
\begin{equation}
 \label{eq:A11}
  A_{11} = 2(\Delta_2\omega^2_2-\Delta_3\omega^2_3)
\end{equation}
\begin{equation}
\label{eq:A12}
 A_{12} = \left(\Delta_1-\Delta_2 + \frac{\Delta_3^2}{I_3} \right)
\end{equation}
\begin{equation}
\label{eq:Deltas}
 \Delta_1 \equiv I_2 - I_3, \: \Delta_2 \equiv I_3 - I_1, \: \Delta_3 \equiv I_1 - I_2,
\end{equation}

From them we obtain the elements $A_{22}$, $A_{33}$, $A_{13}$, and $A_{23}$ by cyclic permutations, with the matrix $A$ being symmetric (Zimmermann 1980).

\section*{Results of Our Numerical Simulations}
\label{Signal}
 \begin{figure}[!ht]
    \centering
    \includegraphics[width=0.9\textwidth]{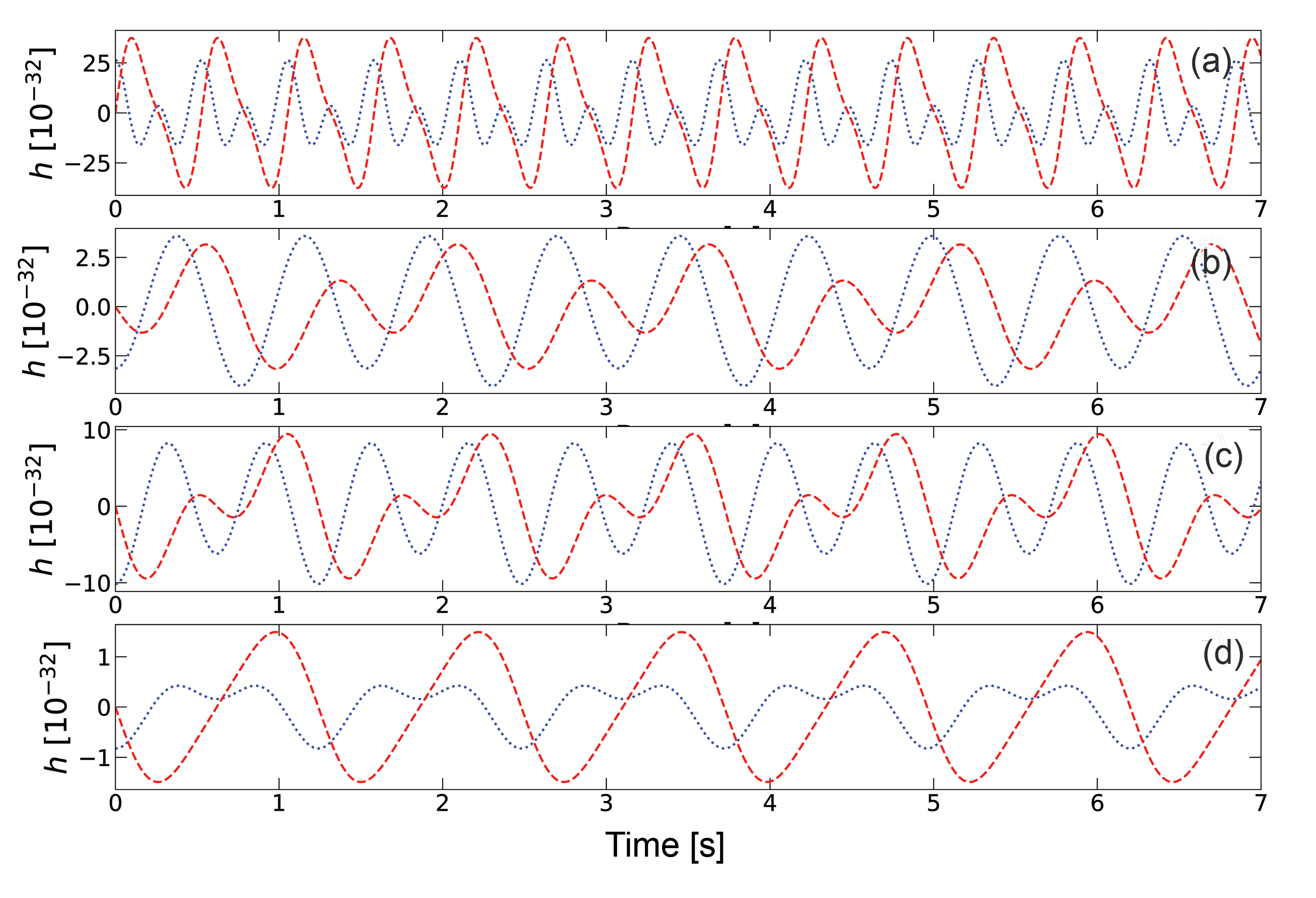}
    \caption{GW profiles as a function of time for $h_{+}$ (dashed line) and $h_{\times}$ (dotted line): (a) $\frac{I_3-I_1}{I_1}=6.7\times10^{-7}$, $\frac{I_2-I_1}{I_1}=2.7\times10^{-7}$, $\theta_0=50^\circ$; (b) $\frac{I_3-I_1}{I_1}=6.7\times10^{-7}$, $\frac{I_2-I_1}{I_1}=2.7\times10^{-7}$, $\theta_0=9.5^\circ$; (c) $\frac{I_1-I_3}{I_3}=\frac{I_2-I_3}{I_3}=6.4\times10^{-7}$, $\theta_0=50^\circ$; (d) $\frac{I_1-I_3}{I_3}=\frac{I_2-I_3}{I_3}=4.2\times10^{-7}$, $\theta_0=11^\circ$.}
    \label{result_fig1}
    \end{figure}
    
\begin{figure}[!ht]
\centering
\includegraphics[width=.75\textwidth]{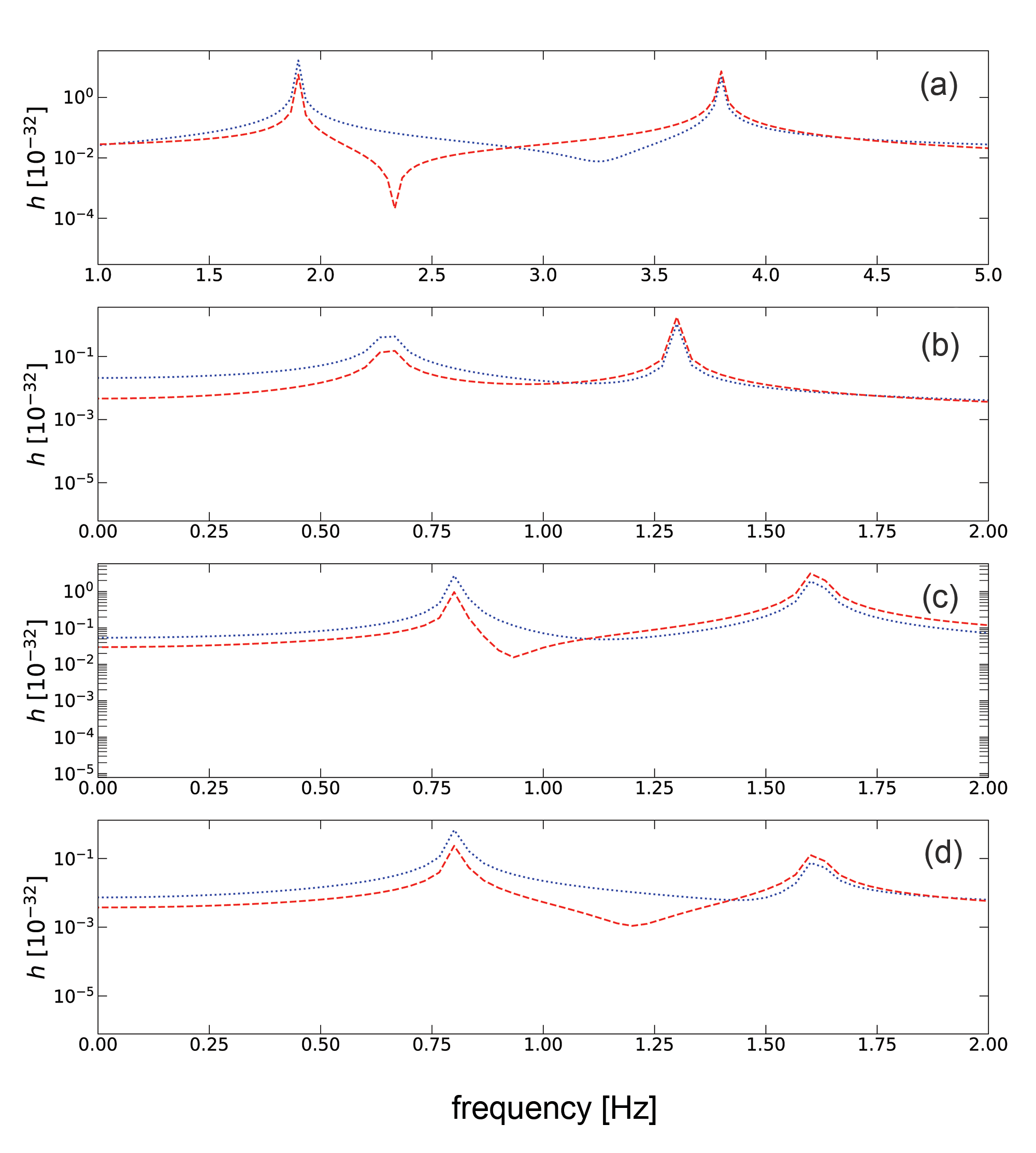}
\caption{Frequency spectrum for $h_{+}$ (dashed line) and $h_{\times}$ (dotted line): (a) $\frac{I_3-I_1}{I_1}=6.7\times10^{-7}$, $\frac{I_2-I_1}{I_1}=2.7\times10^{-7}$, $\theta_0=50^\circ$; (b) $\frac{I_3-I_1}{I_1}=6.7\times10^{-7}$, $\frac{I_2-I_1}{I_1}=2.7\times10^{-7}$, $\theta_0=9.5^\circ$; (c) $\frac{I_1-I_3}{I_3}=\frac{I_2-I_3}{I_3}=6.4\times10^{-7}$, $\theta_0=50^\circ$; (d) $\frac{I_1-I_3}{I_3}=\frac{I_2-I_3}{I_3}=4.2\times10^{-7}$, $\theta_0=11^\circ$.}
\label{result_fig2}
\end{figure}

The variable parameters of our model are the following: $\theta_0$ is the minimum value of the Euler angle $\theta$ and the inertia tensor components $I_1$ and $I_2$. We everywhere take the quantity $I_3$ to be $I_3=10^{45}$~g cm$^2$.
We consider four sets (two for the triaxial ellipsoid and two for the axisymmetric one) of parameters for the NS in Her X-1:
    \begin{itemize}
        \item Model 1: $\frac{I_3-I_1}{I_1}=6.7\times10^{-7}$, $\frac{I_2-I_1}{I_1}=2.7\times10^{-7}$, $\theta_0=50^\circ$ (Kolesnikov et al. 2022).
        \item Model 2: $\frac{I_3-I_1}{I_1}=6.7\times10^{-7}$, $\frac{I_2-I_1}{I_1}=2.7\times10^{-7}$, $\theta_0=9.5^\circ$ (Heyl et al. 2024).
        \item Model 3: $\frac{I_1-I_3}{I_3}=\frac{I_2-I_3}{I_3}=6.4\times10^{-7}$, $\theta_0=50^\circ$ (Postnov et al. 2013).
        \item Model 4: $\frac{I_1-I_3}{I_3}=\frac{I_2-I_3}{I_3}=4.2\times10^{-7}$, $\theta_0=11^\circ$ (Heyl et al. 2024).
    \end{itemize}
    
In addition, in all our models we assume the inclination angle of the NS angular momentum to the line of sight to be $i=110^\circ$ (Heyl et al. 2024).
    
The evolution of the Euler angles and the angular velocities is calculated for the following initial data: $\theta=\theta_0$, $\phi=0$, $\psi=\pi/2$, $\omega_1=a$, $\omega_2=0$, and $\omega_3=b$. In turn, $a$ and $b$ are related to the angle $\theta_0$ and the precession period $T$ via Eqs. (\ref{eq:tgthetha0}) and (\ref{eq:T_prec}):
\begin{equation}
\label{eq:tgthetha0}
 \tg{\theta_0} = \frac{I_1 a}{I_3 b}
\end{equation}
\begin{equation}
\label{eq:T_prec}
 T = \frac{4K(m)}{b} \left[ \frac{I_1 I_2}{(I_3 - I_1)(I_3 - I_2)} \right]^{1/2}
\end{equation}
\begin{equation}
 m=\frac{(I_2-I_1)I_1a^2}{(I_3-I_2)I_3b^2}.
\end{equation}
In Eq. (\ref{eq:T_prec}) $K(m)$ is the complete elliptic integral of the first kind.

The simulated profiles of the GW signal in the two polarizations are presented in Fig.~\ref{result_fig1}. Figure~\ref{result_fig2} contains their frequency spectrum. In our calculations we assumed the distance to the source to be 7~kpc and the NS spin period to be 1.24~s. The signal shape is seen to be sensitive to the values of the parameters. In addition, for the model of an axisymmetric ellipsoid there are two multiple harmonics in the spectrum that depend on nothing but the frequency, while for the nonaxisymmetric models the spectra differ, depending on the angle $\theta_0$.

\section*{The Possibility of Signal Detection}
\label{Detection}
According to Gao et al. (2020), when expanding the GW signal from a precessing ellipsoid in frequencies, two fundamental harmonics must be distinguished at frequencies approximately equal to $f_\mathrm{rot}$ and $2f_\mathrm{rot}$, where $f_\mathrm{rot}$ is the rotation frequency. For Her X-1 $2f_\mathrm{rot}=1.6$~Hz. Such waves can be observed with the Deci-hertz Interferometer Gravitational wave Observatory (DECIGO) telescope and its test version named B-DECIGO that are planned to be launched in the 2030s. The operating range of these two telescopes will be from 0.1 to 10~Hz (Yagi and Seto 2011). Apart from the DECIGO and B-DECIGO missions, there is also the Big Bang Observer (BBO) project of a spaceborne GW detector for operation in this range; its launch date is undetermined even approximately due to its technical complexity. Therefore, in our analysis we will focus on the DECIGO telescope, since it will be more powerful than B-DECIGO and, in contrast to BBO, its launch has already been planned. The main parameter responsible for whether a particular signal can be observed with the GW telescope is its sensitivity.
According to Yagi and Seto (2011), the sensitivity curve for DECIGO is approximated by the expression
 \begin{equation}
 \begin{aligned}
 \label{eq:sensitivity_DECIGO}
  S_n^{DECIGO}(f)=7.05\times10^{-48}\left[1+\left(\frac{f}{\text{7.36~Hz}}\right)^{2}\right]
  +4.08\times10^{-51}\left(\frac{f}{\text{1~Hz}}\right)^{-4}\frac{1}{1+\left(\frac{f}{\text{7.36~Hz}}\right)^2}\\
  +5.33\times10^{-52}\left(\frac{f}{\text{1~Hz}}\right)\text{Hz}^{-1},
  \end{aligned}
 \end{equation}
 
For a cumulative GW signal the signal-to-noise ratio is (Tamanini and Danielski 2019)
\begin{equation}\label{eq:SNR}
  \mathrm{SNR}^2=\frac{2}{S_{n}(f_0)}\sum\limits_{i}\int\limits_0^{T_0}h_i^2(t)dt,
\end{equation}
where $T_0$ is the signal accumulation time, $f_0$ is the signal frequency, $i$ is the number of the channel involved in the signal accumulation, and $h_i$ is the signal amplitude in the $i$th channel. As follows from Yagi and Seto (2011), for DECIGO the number of channels is 2.

Let us estimate how much time it will take for the observations for the signal-to-noise ratio to become 1 in order of magnitude. Let us take an upper limit on the signal semi-amplitude: $\langle h\rangle\approx10^{-31}$. By approximating the integral in (\ref{eq:SNR}) as $\int\limits_0^{T_0}h_i^2(t)dt\approx\langle h\rangle^2T_0$, we find that it will take $\approx1.5\times10^6$~years for the DECIGO telescope to reach a signal-to-noise ratio equal to 1. We added the factor 4 to Eq.~(\ref{eq:SNR}), since the telescope will consist of four clusters of satellites, three satellites in each cluster.

Thus, we see that the signal from Her X-1 is unobservable in the near future.

\section*{Expansion in Terms of the Neutron Star Ellipticity}
\label{analytics}
A typical ellipticity for NSs is $\varepsilon=\frac{I_3-I_1}{I_1}\sim10^{-7}--10^{-6}$. Therefore, $\varepsilon$ is a natural candidate for the solution of the problem of GW modeling based on perturbation theory. The quantity $\delta=\frac{I_2-I_1}{I_3-I_2}$, which shows the degree of deviation of the NS from axial symmetry, acts as an additional parameter.

The Euler equations expressed via the parameters $\varepsilon$ and $\delta$ are rewritten as follows:
\begin{equation}
 \label{eq:euler_params}
 \begin{cases}
  \dot{w_1}  = -\frac{\varepsilon}{(1+\delta)}w_2w_3\\
  \dot{w_2}  = \frac{\varepsilon}{(1+\frac{\delta\varepsilon}{1+\delta})}w_1w_3\\
  \dot{w_3}  = -\frac{\varepsilon\delta}{(1+\delta)(1+\varepsilon)}w_1w_2.
 \end{cases}
\end{equation}
Let us represent the angular velocities as series in $\varepsilon$ ($w_i=\sum\limits_{n=0}^{\infty}w_i^n\varepsilon^n$) and expand the right-hand sides of the Euler equations. We will obtain
\begin{equation}
     \label{eq:euler_seria}
     \begin{cases}
    \dot{w_1}^{n+1}  = -\frac{\sum\limits_{m=0}^n w_2^mw_3^{n-m}}{(1+\delta)}\\
  \dot{w_2}^{n+1}  = -\frac{\sum\limits_{n'=0}^n \sum\limits_{m=0}^{n'}\left(-\frac{\delta}{1+\delta}\right)^{n-n'}w_1^{n'-m}w_3^{m}}{(1+\delta)}\\
  \dot{w_3}^{n+1}  = -\frac{\delta}{1+\delta}\sum\limits_{n'=0}^n \sum\limits_{m=0}^{n'}\left(-1\right)^{n-n'}w_1^{n'-m}w_2^{m}.
  \end{cases}
\end{equation}
The solution of Eqs. (\ref{eq:euler_seria}) for the $n$th order must be a polynomial of degree $n$: $w_i^n=\sum\limits_{k=0}^nw_{i,k}^nt^k$. In the zeroth order the solution will give constant angular velocities $w^0_1$, $w^0_2$, and $w^0_3$. In addition, since we can choose the initial data so that $w_2(t=0)=0$, we have $w_2^0=0$. Therefore, as follows from (\ref{eq:euler_seria}), in the first order in $\varepsilon$ there will be no parameter $\delta$, i.e., the deviation from axial symmetry is an effect of the second order of smallness in ellipticity. Hence, the expressions in the first order in $\varepsilon$ must coincide with the results obtained when expanding the solution for an axisymmetric top into a series in the same parameter. The solution for an axisymmetric top is expressed via elementary functions (see, e.g., Section 36 in the book by Landau and Lifshitz 2004). Substituting the solutions in the form of polynomials into system (\ref{eq:euler_seria}), we find the recurrence relations for $w^{n+1}_{i,k+1}$ via the coefficients in the corresponding polynomials for the lowest orders in $\varepsilon$:
\begin{equation}
\begin{cases}
    w_{1,k+1}^{n+1}  = -\frac{\sum\limits_{j=0}^k\sum\limits_{m=j}^{n-k+j}w_{2,j}^mw_{3,k-j}^{n-m}}{(k+1)(1+\delta)}\\
  w_{2,k+1}^{n+1}  = \frac{1}{k+1}\sum\limits_{n'=k}^n\sum\limits_{j=0}^k \sum\limits_{m=j}^{n'-k+j}\left(-\frac{\delta}{1+\delta}\right)^{n-n'}w_{1,j}^mw_{3,k-j}^{n'-m}\\
  w_{3,k+1}^{n+1}  = -\frac{\delta}{(k+1)(1+\delta)}\sum\limits_{n'=k}^n \sum\limits_{j=0}^k\sum\limits_{m=j}^{n'-k+j}\left(-1\right)^{n-n'}w_{1,j}^mw_{3,k-j}^{n'-m}.
  \end{cases}
  \label{eq:euler_seria_sol}
\end{equation}
It can be made sure through direct calculations that in the first order the expansion of the angular velocities will coincide with the analogous expansion of the solution for an axisymmetric top.

The quaternion components are the next ingredient of the problem. Let us introduce the following notation: we will index the values from 0 to 3 by Greek letters and those from 1 to 3 by Latin letters. The differential equations (\ref{eq:qevolution}) in a compact form are rewritten as
\begin{equation}
    \label{eq:q_evol_compact}
    \dot{q_\mu}=f_{\mu\nu i}w_iq_\nu,
\end{equation}
where on the right-hand side the summation over the repetitive indices $\mu$ and $\nu$ is implied, while $f_{\mu\nu i}$ are the numerical coefficients that are read from the matrix in system (\ref{eq:qevolution}). We will also solve Eqs. (\ref{eq:q_evol_compact}) based on perturbation theory by representing the quaternion components as a series in $\varepsilon$: $q_\mu=\sum\limits_{n=0}^{\infty}q_{\mu}^n\varepsilon^n$. The system of equations for the $n$th order of perturbation theory is written as
\begin{equation}
    \label{eq:q_evol_compact_n}
    \dot{q_\mu}^n=f_{\mu\nu i}w_i^0q_\nu^n+\sum\limits_{n'=1}^nf_{\mu\nu i}w_i^{n'}q_\nu^{n-n'}.
\end{equation}
System (\ref{eq:q_evol_compact_n}) is a linear inhomogeneous system (LIS) with constant coefficients for $q_\mu^n$, with the homogeneous part being specified via $f_{\mu\nu i}w_i^0$ and, in matrix form, being coincident with the matrix from system (\ref{eq:qevolution}). This matrix for the homogeneous part is diagonalized with the eigenvalues $\pm\sqrt{(w_1^0)^2+(w_3^0)^2}$ (given that $w_2^0=0$) located on the diagonal. Denote $w=\sqrt{(w_1^0)^2+(w_3^0)^2}$ and set up the Cauchy problem with the following initial data: $q_0(t=0)=q_{00}$, $q_1(t=0)=q_{11}$, $q_2(t=0)=-q_{11}$, and $q_3(t=0)=q_{00}$, corresponding to the choice of the initial Euler angles $\phi(t=0)=0$ and $\psi(t=0)=\pi/2$. Let us first present the solution of the unperturbed equation for the quaternions, i.e., the solution in the zeroth order in $\varepsilon$:
\begin{equation}
    \label{eq:q_nonperturb}
    \begin{cases}
         q_0^0=
 -\left(q_{10}\frac{w_1}{\omega}+q_{00}\frac{\omega_3}{\omega}\right)\sin\left(\frac{\omega t}{2}\right)+q_{00}\cos\left(\frac{\omega t}{2}\right)\\
 q_1^0=\left(q_{00}\frac{w_1}{w}-q_{10}\frac{w_3}{w}\right)\sin\left(\frac{wt}{2}\right)+q_{10}\cos\left(\frac{wt}{2}\right)\\
 q_2^0=-q_{10}\cos\left(\frac{wt}{2}\right)+\left(q_{10}\frac{w_3}{w}-q_{00}\frac{w_1}{w}\right)\sin\left(\frac{wt}{2}\right)\\
 q_3^0=
 q_{00}\cos\left(\frac{wt}{2}\right)+\left(q_{00}\frac{w_3}{w}+q_{10}\frac{w_1}{w}\right)\sin\left(\frac{wt}{2}\right).
    \end{cases}
\end{equation}
The equations for all the succeeding orders of perturbation theory will be linear inhomogeneous equations with a resonance of multiplicity 1. Its general solution can be found as a sum of the general solution of the linear homogeneous system (LHS) and the particular solution of the LIS. By induction, it can be shown that the particular solution of the LIS for order $n$ will be presented in the form
\begin{equation}
    \label{eq:q_p_seria_sol}
    q_\mu^{n,\mathrm{p}}=\cos\left(\frac{wt}{2}\right)\sum\limits_{k=0}^{2n}q^{n,\mathrm{p},\mathrm{c}}_{\mu,k}t^k+\sin\left(\frac{wt}{2}\right)\sum\limits_{k=0}^{2n}q^{n,\mathrm{p,s}}_{\mu,k}t^k
\end{equation}
where $q^{n,\mathrm{p,s}}_{\mu,k}$ and $q^{n,\mathrm{p,c}}_{\mu,k}$ are the numerical coefficients. These coefficients can be found by solving the system of linear algebraic equations obtained by substituting the solution (\ref{eq:q_p_seria_sol}) into system (\ref{eq:q_evol_compact_n}):
\begin{equation}
    \label{eq:slau}
    \begin{cases}
        q^{n,\mathrm{p,s}}_{\mu,2n}\frac{w}{2}-f_{\mu\nu i}w_i^0q^{n,\mathrm{p,c}}_{\nu,2n}=0\\
        q^{n,\mathrm{p,c}}_{\mu,2n}\frac{w}{2}+f_{\mu\nu i}w_i^0q^{n,\mathrm{p,s}}_{\nu,2n}=0\\
        q^{n,\mathrm{p,s}}_{\mu,k}\frac{w}{2}+q^{n,\mathrm{p,c}}_{\mu,k+1}(k+1)-f_{\mu\nu i}w_i^0q^{n,\mathrm{p,c}}_{\nu,k}
        =f_{\mu\nu i}\sum\limits_{j=1}^k\sum\limits_{n'=k//2}^{n-j}q^{n,\mathrm{p,c}}_{\nu,k-j}w^{n-n'}_{i,j},\quad 0\leq k<2n\\
        -q^{n,\mathrm{p,c}}_{\mu,k}\frac{w}{2}+q^{n,\mathrm{p,s}}_{\mu,k+1}(k+1)-f_{\mu\nu i}w_i^0q^{n,\mathrm{p,s}}_{\nu,k} 
        =f_{\mu\nu i}\sum\limits_{j=1}^k\sum\limits_{n'=k//2}^{n-j}q^{n,\mathrm{p,s}}_{\nu,k-j}w^{n-n'}_{i,j},\quad 0\leq k<2n.
    \end{cases}
\end{equation}
Here and below, we define $k//2$ as follows:
\begin{equation}
    \label{eq:div_def}
    \begin{cases}
        k//2=k/2,\quad k\in{0,2,4,6,...}\\
        k//2=(k+1)/2,\quad k\in{1,3,5,7,...} .
    \end{cases}
\end{equation}
Unfortunately, we failed to find the solution of Eqs. (\ref{eq:slau}) in a closed form in the case of an arbitrary order $n$. However, for each specific order $n$ it can be solved, for example, by the Gaussian method. It should be noted that since there is a resonance of multiplicity 1 in the original LIS, not all of the equations written above are algebraically independent. Two equations will be dependent on all of the remaining ones, leaving arbitrary free coefficients when solving the system.

When the particular solution of the inhomogeneous system has been constructed, i.e., the coefficients $q^{n,\mathrm{p,s}}_{\mu,k}$ and $q^{n,\mathrm{p,c}}_{\mu,k}$ have been found, the next step is to find the solution of the Cauchy problem with the initial data $q_\mu^n(t=0)=0$ ($n>0$). We set the conditions based on the fact that $q_\mu^0$ satisfies the initial data following from the initial data for the Euler angles. The general solution of the linear homogeneous system is written as
\begin{equation} \label{eq:q_u}
    \begin{cases}
        q_0^\mathrm{u}=C_1\frac{w_1^0}{w_3}\cos\left(\frac{wt}{2}\right)+C_2\frac{w_1^0}{w_3^0}\sin\left(\frac{wt}{2}\right)-C_3\frac{w}{w_3^0}\sin\left(\frac{wt}{2}\right)+C_4\frac{w}{w_3^0}\cos\left(\frac{wt}{2}\right)\\
        q_1^\mathrm{u}=C_1\frac{w}{w_3^0}\sin\left(\frac{wt}{2}\right)-C_2\frac{w}{w_3^0}\cos\left(\frac{wt}{2}\right)+C_3\frac{w_1^0}{w_3^0}\cos\left(\frac{wt}{2}\right)+C_4\frac{w_1^0}{w_3^0}\sin\left(\frac{wt}{2}\right)\\
        q_2^\mathrm{u}=C_1\cos\left(\frac{wt}{2}\right)+C_2\sin\left(\frac{wt}{2}\right)\\
        q_3^\mathrm{u}=C_3\cos\left(\frac{wt}{2}\right)+C_4\sin\left(\frac{wt}{2}\right),
    \end{cases}    
\end{equation}
where $C_{1\ldots4}$ are the constants of integration. In order $n$ in $\varepsilon$ the components $q_\mu^n$ are expressed as
\begin{equation}
    \label{eq:q_seria_sol}
    q_\mu^{n}=\cos\left(\frac{wt}{2}\right)\sum\limits_{k=0}^{2n}q^{n,\mathrm{c}}_{\mu,k}t^k+\sin\left(\frac{wt}{2}\right)\sum\limits_{k=0}^{2n}q^{n,\mathrm{s}}_{\mu,k}t^k,
\end{equation} 
where $q^{n,\mathrm{c}}_{\mu,0}=0$, $q^{n,\mathrm{s}}_{0,0}=q^{n,\mathrm{p,s}}_{0,0}+q^{n,\mathrm{p,c}}_{3,0}w_3^0/w$, $q^{n,\mathrm{s}}_{1,0}=q^{n,\mathrm{p,s}}_{1,0}-q^{n,\mathrm{p,c}}_{2,0}w_3^0/w$, $q^{n,\mathrm{s}}_{2,0}=q^{n,\mathrm{p,s}}_{2,0}-q^{n,\mathrm{p,c}}_{3,0}w_1^0/w$, $q^{n,\mathrm{s}}_{3,0}=q^{n,\mathrm{p,s}}_{3,0}+q^{n,\mathrm{p,c}}_{2,0}w_1^0/w$, while the following is true for the remaining coefficients: $q^{n,\mathrm{c}}_{\mu,k}=q^{n,\mathrm{p,c}}_{\mu,k}$ and $q^{n,\mathrm{s}}_{\mu,k}=q^{n,\mathrm{p,s}}_{\mu,k}$.

Next, it is necessary to present the expansion of the rotation matrix $\mathcal{R}_{ij}$ into a series in $\varepsilon$: $\mathcal{R}_{ij}=\sum\limits_{n=0}^{\infty}\mathcal{R}^n_{ij}\varepsilon^n$.
The structure of the element $\mathcal{R}_{ij}$ is $\mathcal{R}_{ij}=g_{ij\mu\nu}q_\mu q_\nu$, where the summation is over the repetitive indices $\mu,\nu$, while $g_{ij\mu\nu}$ are the constant coefficients that are read from the matrix (\ref{eq:R_matrix}). As a result of some calculations, we obtain
\begin{equation}
    \label{eq:R_n_pol}
    \mathcal{R}_{ij}^n=\frac{\sin(wt)}{2}\sum\limits_{k=0}^{2n}\mathcal{R}_{ij,k}^{n,\mathrm{s}}t^k 
    + \frac{\cos(wt)}{2}\sum\limits_{k=0}^{2n}\mathcal{R}_{ij,k}^{n,\mathrm{c}}t^k +\sum\limits_{k=0}^{2n}\mathcal{R}_{ij,k}^{n,\mathrm{f}}t^k,
\end{equation}
where the coefficients $\mathcal{R}_{ij,k}^{n,\mathrm{s}}$, $\mathcal{R}_{ij,k}^{n,\mathrm{c}}$, and $\mathcal{R}_{ij,k}^{n,\mathrm{f}}$ are expressed as follows:
\begin{equation}
    \label{eq:R_ij_n_coef}    
        \begin{aligned}
\mathcal{R}_{ij,k}^{n,\mathrm{c}}=g_{ij\mu\nu}\sum\limits_{n'=k//2}^{n}\sum_{j'=0}^k\left(q_{\mu,j'}^{n',\mathrm{c}}q_{\nu,k-j'}^{n-n',\mathrm{c}}-q_{\mu,j'}^{n',\mathrm{s}}q_{\nu,k-j'}^{n-n',\mathrm{s}}\right)\\
\mathcal{R}_{ij,k}^{n,\mathrm{s}}=g_{ij\mu\nu}\sum\limits_{n'=k//2}^{n}\sum_{j'=0}^k\left(q_{\mu,j'}^{n',\mathrm{c}}q_{\nu,k-j'}^{n-n',\mathrm{s}}+q_{\mu,j'}^{n',\mathrm{s}}q_{\nu,k-j'}^{n-n',\mathrm{c}}\right)\\
    \mathcal{R}_{ij,k}^{n,f}=\frac{g_{ij\mu\nu}}{2}\sum\limits_{n'=k//2}^{n}\sum_{j'=0}^k\left(q_{\mu,j'}^{n',\mathrm{c}}q_{\nu,k-j'}^{n-n',\mathrm{c}}+q_{\mu,j'}^{n',\mathrm{s}}q_{\nu,k-j'}^{n-n',\mathrm{s}}\right),
    \end{aligned}
\end{equation}
the summation over the repetitive indices $\mu,\nu$ is implied.

Next, let us represent the quantities $\Delta_i$ as series in $\varepsilon$ using their definition:
\begin{equation}
    \label{eq:Delta_n}
    \begin{cases}
        \Delta_1=I_2-I_3=\frac{I_3}{1+\delta}\sum\limits_{n=1}^\infty(-1)^n\varepsilon^n\\
        \Delta_2=I_3-I_1=I_3\sum\limits_{n=1}^{\infty}(-1)^{n-1}\varepsilon^n\\
        \Delta_3=\frac{\delta I_3}{1+\delta}\sum\limits_{n=1}^{\infty}(-1)^n\varepsilon^n.
    \end{cases}
\end{equation}
Thereafter, let us present the components $A_{ij}$ in the form of series: $A_{ij}=\varepsilon\sum\limits_{n=0}^\infty A_{ij}^n\varepsilon^n$. Here, one factor $\varepsilon$ was specially taken outside the summation sign. It arises because the series for $\Delta_k$ begin from the first term. Since only the products of $\Delta_k$ and the angular velocities, which, in turn, are expanded into series in $\varepsilon$ with the coefficients of the series representable as polynomials in $t$, enter into the expression for $A_{ij}$, the coefficients $A_{ij}^n$ can also be written as polynomials in $t$: $A_{ij}^{n+1}=\sum\limits_{k=0}^{n}A_{ij,k}^{n+1}t^k$. The expressions for the diagonal components are
\begin{equation}\label{eq:A_n_pl_one_diag}
 A_{11,k}^{n+1}=2I_3\left(\sum\limits_{n'=k}^n(-1)^{n-n'}\sum\limits_{j=0}^k\sum\limits_{n''=j}^{n'-k+j}w_{2,j}^{n''}w_{2,k-j}^{n'-n''} - \frac{\delta}{1+\delta}\sum\limits_{n'=k}^n(-1)^{n-n'+1}\sum\limits_{j=0}^k\sum\limits_{n''=j}^{n'-k+j}w_{3,j}^{n''}w_{3,k-j}^{n'-n''}\right)\\
\end{equation}
\begin{equation}
    A_{22,k}^{n+1}=\frac{2I_3}{1+\delta}\left(\delta\sum\limits_{n'=k}^n(-1)^{n-n'+1}\sum\limits_{j=0}^k 
    \sum\limits_{n''=j}^{n'-k+j}w_{3,j}^{n''}w_{3,k-j}^{n'-n''}- \sum\limits_{n'=k}^n(-1)^{n-n'+1}\sum\limits_{j=0}^k\sum\limits_{n''=j}^{n'-k+j}w_{1,j}^{n''}w_{1,k-j}^{n'-n''}\right)\\
\end{equation}
\begin{equation}
    A_{33,k}^{n+1}=2I_3\left(\frac{1}{1+\delta}\sum\limits_{n'=k}^n(-1)^{n-n'+1}\sum\limits_{j=0}^k \sum\limits_{n''=j}^{n'-k+j}w_{1,j}^{n''}w_{1,k-j}^{n'-n''}--\sum\limits_{n'=k}^n(-1)^{n-n'}\sum\limits_{j=0}^k\sum\limits_{n''=j}^{n'-k+j}w_{2,j}^{n''}w_{2,k-j}^{n'-n''}\right)\\
\end{equation}
\begin{equation}
    A_{12,k}^{n+1}=I_3\left(\frac{2+\delta}{1+\delta}\sum\limits_{n'=k}^n(-1)^{n-n'+1}\sum\limits_{j=0}^k\sum\limits_{n''=j}^{n'-k+j}w_{1,j}^{n''}w_{2,k-j}^{n'-n''}+D_{3,k}^{n+1}\right)\\
\end{equation}
\begin{equation}
    A_{13,k}^{n+1}=I_3\left(\frac{1-\delta}{1+\delta}\sum\limits_{n'=k}^n(-1)^{n-n'+1}\sum\limits_{j=0}^k\sum\limits_{n''=j}^{n'-k+j}w_{1,j}^{n''}w_{3,k-j}^{n'-n''} +D_{2,k}^{n+1}\right)\\
\end{equation}
\begin{equation}
    A_{23,k}^{n+1}=I_3\left(\frac{1+2\delta}{1+\delta}\sum\limits_{n'=k}^n(-1)^{n-n'}\sum\limits_{j=0}^k\sum\limits_{n''=j}^{n'-k+j}w_{2,j}^{n''}w_{3,k-j}^{n'-n''} +D_{1,k}^{n+1}\right)\\
\end{equation}
\begin{equation}
    A_{21,k}^{n+1}=A_{12,k}^{n+1},\quad A_{32,k}^{n+1}=A_{23,k}^{n+1},\quad
    A_{31,k}^{n+1}=A_{13,k}^{n+1}
%\end{cases}
\end{equation}

%Внедиагональные компоненты выражаются как:
%\begin{multline}
%    \begin{cases}
%        A_{12,k}^{n+1}=I_3\left(\frac{2+\delta}{1+\delta}\sum\limits_{n'=k}^n(-1)^{n-n'+1}\sum\limits_{j=0}^k\sum\limits_{n''=j}^{n'-k+j}w_{1,j}^{n''}w_{2,k-j}^{n'-n''}\right.\\\left.+D_{3,k}^{n+1}\right)\\
%    A_{13,k}^{n+1}=I_3\left(\frac{1-\delta}{1+\delta}\sum\limits_{n'=k}^n(-1)^{n-n'+1}\sum\limits_{j=0}^k\sum\limits_{n''=j}^{n'-k+j}w_{1,j}^{n''}w_{3,k-j}^{n'-n''}\right.\\\left.+D_{2,k}^{n+1}\right)\\
%    A_{23,k}^{n+1}=I_3\left(\frac{1+2\delta}{1+\delta}\sum\limits_{n'=k}^n(-1)^{n-n'}\sum\limits_{j=0}^k\sum\limits_{n''=j}^{n'-k+j}w_{2,j}^{n''}w_{3,k-j}^{n'-n''}\right.\\\left.+D_{1,k}^{n+1}\right)\\
%    A_{21,k}^{n+1}=A_{12,k}^{n+1},\quad A_{32,k}^{n+1}=A_{23,k}^{n+1},\\ A_{31,k}^{n+1}=A_{13,k}^{n+1}
%    \end{cases}
%    \label{eq:A_mat_nondiag}
%\end{multline}
Here, $D_1^{0}=D_1^{1}=D_2^{0}=D_2^{1}=D_3^{0}=D_3^{1}=0$ and
\begin{equation}
    \label{eq:D_n}
    \begin{cases}
        D_{3,k}^{n+2}=\frac{\delta^2}{(1+\delta)^2}\sum\limits_{n'=k}^n\left(\sum\limits_{n''=0}^{n-n'}(-1)^{n-n'}\right) \left(\sum\limits_{j=0}^{k}\sum\limits_{n''=j}^{n'-k+j}w_{1,j}^{n''}w_{2,k-j}^{n'-n''}\right)\\
        D_{2,k}^{n+2}=\sum\limits_{n'=k}^n\left(\sum\limits_{n''=0}^{n-n'}(-1)^{n-n'}\right) \left(\sum\limits_{j=0}^{k}\sum\limits_{n''=j}^{n'-k+j}w_{1,j}^{n''}w_{3,k-j}^{n'-n''}\right)\\
        D_{1,k}^{n+2}=\frac{1}{(1+\delta)^2}\sum\limits_{n'=k}^n\left(\sum\limits_{n''=0}^{n-n'}(-1)^{n-n'}\right) \left(\sum\limits_{j=0}^{k}\sum\limits_{n''=j}^{n'-k+j}w_{2,j}^{n''}w_{3,k-j}^{n'-n''}\right)
    \end{cases}
\end{equation}

When the expansions into series in $\varepsilon$ have been found for the matrices $\mathcal{R}_{ij}$ and $A_{ij}$, we can write the expression for the GW amplitudes also as a series in variable $\varepsilon$. Let $\lambda$ index the GW polarizations: $h_\lambda\in\{h_{+},h_{\times}\}$. The general structure for $h_\lambda$ is $h_\lambda=\mathcal{R}_{ik}\mathcal{R}_{jl}A_{kl}\tilde{g}^\lambda_{ij}$, where the summation over all repetitive indices is implied and $\tilde{g}^\lambda_{ij}$ are the numerical coefficients that are related to the angle $i$ between the observer's line of sight and the angular momentum vector and are read from Eqs. (\ref{eq:h_pl}) and (\ref{eq:h_cr}). We will seek an expansion for $h_\lambda$ in the form
\begin{equation}
    h_\lambda=\varepsilon\sum\limits_{n=0}^\infty h_\lambda^n\varepsilon^n
\end{equation}
Note that in this case one factor $\varepsilon$ is outside the summation sign, i.e., the summation begins from the first order in this variable. This follows from the fact that in the expansion for $A_{kl}$ the summation starts from the first order. Hence it follows that a rotating spherically symmetric NS ($\varepsilon=0$) does not emit any GWs. In addition, as has already been said above, the deviation from axial symmetry $\delta$ arises only in the next order in $\varepsilon$, i.e., the leading contribution will coincide with that from an axisymmetric top.

Since the product $\mathcal{R}_{ik}\mathcal{R}_{jl}$ enters into the expression for $h_\lambda$, the coefficient $h^n_\lambda$ will have the following structure:
\begin{multline}
h^n_\lambda=\cos(2wt)\sum_{k'=0}^{2n}h^{n,\mathrm{2c}}_{\lambda,k'}t^{k'}+\sin(2wt)\sum_{k'=0}^{2n}h^{n,\mathrm{2s}}_{\lambda,k'}t^{k'}
+\cos(wt)\sum_{k'=0}^{2n}h^{n,\mathrm{c}}_{\lambda,k'}t^{k'}+\sin(wt)\sum_{k'=0}^{2n}h^{n,\mathrm{s}}_{\lambda,k'}t^{k'} +\sum_{k'=0}^{2n}h^{n,\mathrm{f}}_{\lambda,k'}t^{k'}.
    \label{eq:h_n}
\end{multline}
The contributions from two harmonics, $2wt$ and $wt$, will be present in this expression. Both harmonics will not disappear even in the first order of smallness in $\varepsilon$, consistent with the non-perturbative analysis performed by Zimmermann and Szedenits (1979). The coefficients of these harmonics will be polynomials in $t$ of degree $2n$. Let us present the expressions for the numerical coefficients of these polynomials:
\begin{equation}
h^{n,\mathrm{2c}}_{\lambda,k'}=\frac{\tilde{g}_{ij}}{2}\sum\limits_{j'=0}^{k'}\sum_{n'=k'//2}^{n+1-(k'-j')//2}A_{kl}^{n+1-n'}  \sum\limits_{j''=0}^{j'}\sum\limits_{n''=j'//2}^{n'-(j'-j'')//2}\left(\mathcal{R}_{ik,j''}^{n',\mathrm{c}}\mathcal{R}_{jl,j'-j''}^{n'-n'',\mathrm{c}}-\mathcal{R}_{ik,j''}^{n',\mathrm{s}}\mathcal{R}_{jl,j'-j''}^{n'-n'',\mathrm{s}}\right)\\
\end{equation}
\begin{equation}
h^{n,\mathrm{2s}}_{\lambda,k'}=\frac{\tilde{g}_{ij}}{2}\sum\limits_{j'=0}^{k'}\sum_{n'=k'//2}^{n+1-(k'-j')//2}A_{kl}^{n+1-n'} \sum\limits_{j''=0}^{j'}\sum\limits_{n''=j'//2}^{n'-(j'-j'')//2}\left(\mathcal{R}_{ik,j''}^{n',\mathrm{c}}\mathcal{R}_{jl,j'-j''}^{n'-n'',\mathrm{s}}+\mathcal{R}_{ik,j''}^{n',\mathrm{s}}\mathcal{R}_{jl,j'-j''}^{n'-n'',\mathrm{c}}\right)\\
\end{equation}
\begin{equation}
h^{n,\mathrm{c}}_{\lambda,k'}=\tilde{g}_{ij}\sum\limits_{j'=0}^{k'}\sum_{n'=k'//2}^{n+1-(k'-j')//2}A_{kl}^{n+1-n'} \sum\limits_{j''=0}^{j'}\sum\limits_{n''=j'//2}^{n'-(j'-j'')//2}\left(\mathcal{R}_{ik,j''}^{n',\mathrm{c}}\mathcal{R}_{jl,j'-j''}^{n'-n'',\mathrm{f}}+\mathcal{R}_{ik,j''}^{n',\mathrm{f}}\mathcal{R}_{jl,j'-j''}^{n'-n'',\mathrm{c}}\right)\\
\end{equation}
\begin{equation}
h^{n,\mathrm{s}}_{\lambda,k'}=\tilde{g}_{ij}\sum\limits_{j'=0}^{k'}\sum_{n'=k'//2}^{n+1-(k'-j')//2}A_{kl}^{n+1-n'} \sum\limits_{j''=0}^{j'}\sum\limits_{n''=j'//2}^{n'-(j'-j'')//2}\left(\mathcal{R}_{ik,j''}^{n',\mathrm{s}}\mathcal{R}_{jl,j'-j''}^{n'-n'',\mathrm{f}}+\mathcal{R}_{ik,j''}^{n',\mathrm{f}}\mathcal{R}_{jl,j'-j''}^{n'-n'',\mathrm{s}}\right)
\end{equation}
\begin{equation}
    h^{n,\mathrm{f}}_{\lambda,k'}=\frac{\tilde{g}_{ij}}{2}\sum\limits_{j'=0}^{k'}\sum_{n'=k'//2}^{n+1-(k'-j')//2}A_{kl}^{n+1-n'} \sum\limits_{j''=0}^{j'}\sum\limits_{n''=j'//2}^{n'-(j'-j'')//2}\left(\mathcal{R}_{ik,j''}^{n',\mathrm{c}}\mathcal{R}_{jl,j'-j''}^{n'-n'',\mathrm{c}}+\mathcal{R}_{ik,j''}^{n',\mathrm{s}}\mathcal{R}_{jl,j'-j''}^{n'-n'',\mathrm{s}}\right. \left.+2\mathcal{R}_{ik,j''}^{n',\mathrm{f}}\mathcal{R}_{jl,j'-j''}^{n'-n'',\mathrm{f}}\right)\\
    \label{eq:h_n_k}
\end{equation}

In this section we successively deduced the analytical recurrence formulas for the coefficients of the expansion in $\varepsilon$ for all of the quantities important for the construction of the GW front. Unfortunately, we failed to obtain them in a closed form, since this requires solving the system of linear equations (\ref{eq:div_def}), for the components of which we failed to select the general formula. In contrast to the picture of a purely oscillatory process, in which the GW frequency is twice the frequency of the system, the GW signal from a nonspherical precessing body contains two harmonics with frequencies $w$ and $2w$ and an anharmonic part. It is also clearly seen that the rotating body must be nonspherical for the generation of GWs, while the deviation from axial symmetry makes a contribution from the second order of smallness in $\varepsilon$.

Note that, as far as we know, these expressions are the first attempt to solve the problem of GW front modeling by the analytical method of perturbations in $\varepsilon$. Previously, Zimmermann (1980), Van Den Broeck (2005), and Gao et al. (2020) derived the analytical expressions through the expansion into series in small parameters $\delta$ and $\gamma=\tan(\theta_\mathrm{min})$. The choice of these parameters allows one to present compact expressions, which we failed to achieve. However, these parameters per se for NSs may not be small. For example, in the Her X-1 precession models considered in this paper the parameter $\gamma$ takes values of 0.2 and 1.2, while $\delta$ for the nonaxisymmetric case is 0.7. Thus, the parameter $\varepsilon$ is a more suitable candidate for constructing perturbation theory. Our formulas are applicable not only to the GWs from the pulsar Her X-1, but also to other NSs, for example, the millisecond pulsar Sco X-1, which is one of the candidates for the search of continuous GWs with ground-based detectors of the second and third generations (see, e.g., Riles 2023).

\section*{Conclusions}
\label{Conclusion}

Our paper is devoted to studying the GW signal produced by the NS precession in the pulsar Her X-1. The signal profile was constructed in the quadrupole approximation and depends strongly on the precession parameters with regard to which there are disagreements in the currently available works. Based on the derived amplitude of GWs, we estimate their potential observability. We established that although the GW frequencies fall into the range of DECIGO, B-DECIGO, and BBO observations, none of these telescopes can detect the GW signal.

In this paper we did not consider the effects associated with the nonrigidity of the NS rotation, including the possibility of the rotation of a superfluid component around an axis that is misaligned with the stellar rotation axis (Jones 2010). We also ignored the NS eigenmodes; in particular, it is known that the r-modes can lead to the emission of GWs at frequencies comparable to the rotation frequency (see, e.g., Riles 2023), which potentially is able to affect the overall signal shape. In addition, the change in the precession period due to accretion was ignored in the model, whereas, according to the estimates of Postnov et al. (2013), the corresponding rate of change is $\dot{T}=4\times10^{-5}$. Allowance for the listed effects can lead to a refinement of the quantitative signal characteristics.

The amplitude of GWs from a precessing ellipsoid is proportional to the square of the rotation frequency and inversely proportional to the distance to the source. Her X-1 has one of the shortest NS spin periods among the X-ray pulsars (see, e.g., Bildsten et al. 1997) and is one of the best-studied sources of this type. The fact that the GWs from the NS precession in Her X-1 remain unobservable even in the presence of independent constraints on the precession parameters from X-ray data suggests that at the current detector sensitivity level they are extremely difficult to detect. In the foreseeable future, apparently, X-ray pulsars of this type can hardly be expected to serve as promising sources for GW observations.

Apart from the calculations for the pulsar Her X-1 based on the numerical solution of the problem of the rotation of an absolutely rigid body, we constructed the GW profile through an expansion in terms of the small NS ellipticity parameter. We derived the recurrence relations that allow one to determine the profile of the GW signal from precessing NSs in an arbitrary order of perturbation theory. These formulas can be used to speed up the calculations in problems related to the search for GWs from precessing NSs with millisecond periods that, as expected, can be detected already in the near future.

The work of I.D. Markozov was supported by grant no.~075-15-2024-647 of the Ministry of Education and Science of the Russian Federation. We are grateful to the anonymous referee for a number of useful remarks to improve the paper.

\end{document}